\documentclass[12pt]{article}
\pdfoutput=1
\usepackage{epsfig,graphicx}
\begin{document}
\title{Tumbling of a rigid rod in a shear flow}

\author{J.~M.~J. van Leeuwen and  H.~W.~J. Bl\"ote  \\
Instituut-Lorentz,  Universiteit Leiden, \\
Niels Bohrweg 2, 2333 CA Leiden, The   Netherlands}

\maketitle

\begin{abstract}
The tumbling of a rigid rod in a shear flow is analyzed in the high viscosity limit.
Following Burgers, the Master Equation is derived for the probability distribution of the orientation
of the rod. The equation contains one dimensionless number, the Weissenberg number,
which is the ratio of the shear rate and the orientational diffusion constant.
The equation is solved for the stationary state distribution for arbitrary 
Weissenberg numbers, in particular for the limit of high Weissenberg numbers.
The stationary state gives an interesting flow pattern for the orientation of the rod,
showing the interplay between flow due to the driving shear force and diffusion due
to the random thermal forces of the fluid. The average tumbling time and tumbling
frequency are calculated as a function of the Weissenberg number. A simple cross-over
function is proposed which covers the whole regime from small to large Weissenberg 
numbers.
\end{abstract} 

keywords: shear flow, polymers, Fokker-Planck equation.

\section{Introduction}

The behavior of small particles immersed in a shear flow has been of interest for a long time. 
On the one hand the particles influence the flow properties and on the other hand the flow
controls the motion of the particles. In biological systems with flows through narrow
channels, one frequently encounters a sheared flow-field carrying polymeric particles.
In recent experiments \cite{gerashchenko, kanstler} one has focused the distribution 
of the particles as a function of their orientation. Particularly interesting are the experiments of 
Harasim et al. \cite{bausch} where the motion of fragments of f-actin in a shear flow has been 
recorded such that the tumbling of the fragments can be seen {\it ad oculos}. 
The motion of the particles is quasi-periodic, with a stochastically distributed  period.
This makes the explanation also theoretically  of much interest,  as it is a combined effect 
of systematic as well as thermal forces of the flow on the particles. Several theoretical
papers \cite{winckler, das, abreu, winckler2} haved been  devoted to the analysis of the 
motion using various approximations, but it turns out to be difficult to theoretically 
extract the periodicity.
For a recent survey of the motion of solid objects in a flow see \cite{dhont}.

In this recent literature no mention is made of a fundamental contribution of J.M. Burgers
\cite{burgers}, who considered the simplest version of the problem: 
that of a rigid object in a shear flow.\footnote{Indeed the study of Burgers of 1938 
is not easily accessible as it is not available online. The authors are indebted to H.~N.~W. 
Lekkerkerker for making this paper available to them.}
Burgers derived the equation for the probabilty distribution for the steady state and 
analyzed the solution for the first few orders in an expansion in powers of the Weissenberg
number, which is a dimensionless measure of the strength of the shear force.
In this paper we revisit the problem that Burgers discussed and extend the solution
to arbitrary values of the Weissenberg number $W$ by numerical exact methods.
In particular we are interested in the large-$W$ limit.
This allows us to derive the flow pattern in orientation space, to extract the periodicity
and to  give the average tumbling time.
 
In this note  we consider, for simplicity, a rigid rod consisting of a number of stiffly aligned
monomers. Polymers may well be approximated by this model if their length is shorter than 
the persistence length. One can form a simple Hamiltonian model of interacting beads 
\cite{gerard} and tune the parameters of the model such that they accurately reproduce the
force-extension curve of a class of polymers \cite{wang}. Indeed, as the experiments of \cite{bausch}
show, a polymer like f-actin tumbles almost without bending when the length
is shorter than the persistence length \cite{liu}.

The influence of the fluid on the rod is given by the Langevin equation in the
high viscosity limit, leaving out acceleration effects. We start by deriving the Langevin equation 
for a rod of rigidly aligned monomers and transform it into the equivalent Fokker-Planck equation, 
which was Burgers' line of approach.  Then we discuss the expansion of the solution in powers of
$W$ and indicate that the convergence radius of the series is of the order $W \simeq 1$. As a 
preparation for and illustration of the spherical geometry we show how the 
problem confined to a circular geometry admits an exact analytical solution, which
demonstrates scaling behavior in the large-$W$ limit. We expand the spherical solution
for arbitrary Weissenberg numbers $W$ in terms of adapted basis functions and
solve the partial differential equation by an optimization process. Using this optimization we
obtain an accurate solution for Weissenberg numbers up to $W=30$.
With the solution we determine the flow pattern in  the orientation space, yielding the average 
period of the tumbling and the average tumbling frequency. 
Finally we analyze the scaling solution for the large-$W$ limit and give 
the properties of the tumbling process in this limit. The scaling limit matches perfectly
with the results for $W \simeq 30$ and an simple interpolation formula is given, covering the 
whole range of Weissenberg numbers.

\section{The Langevin Equations for the rod}

In the high viscosity limit the Langevin equation the equation of motion for the monomers,
kicked around by random forces and slowed down by friction,  reads
\begin{equation} \label{a1}
\frac{d {\bf r}_n}{ dt} = - {1 \over \xi}{\partial {\cal H} \over \partial {\bf r}_n} +
\dot{\gamma} \, (y_n -Y_{cm}) \, \hat{\bf x} +{\bf g}_n .
\end{equation}
Here $\cal H$ is the hamiltonian of the monomers and $\xi$ is the friction coefficient. 
The first term on the right hand side of the equation represents the internal forces, keeping
the monomers aligned and equidistant. The second term is the shear force with flow  in 
$\hat{x}$ direction with a gradient in the $\hat{y}$ direction.
$\dot{\gamma}$ is the shear rate. 
$Y_{cm}$ is the $y$ coordinate of the center-of-mass of the chain and $y_n$ is the $y$ 
coordinate of monomer $n$. From now on we subtract the center-of mass motion, 
by taking the positions ${\bf r}_n$ with respect to the center-of-mass.
In total we have $N+1$ monomers in the chain. The last term in (\ref{a1}) gives the 
influence of the random force ${\bf g}_n$, which has the correlation function
\begin{equation} \label{a2}
\langle g^\alpha_m (t) \,  g^\beta_n (t' ) \rangle = (2 \, k_B \, T /\xi)  \, \delta^{\alpha,\beta} \, 
\delta_{m,n} \delta (t-t').
\end{equation} 

By taking the outer product of each equation with ${\bf r}_n$, we get on the left hand side 
the instantaneous total angular momentum $\bf L$ of the rod
\begin{equation} \label{a3}
{\bf L} = \sum_n {\bf r}_n \times {d {\bf r}_n \over dt}.  
\end{equation} 
On the right hand side the terms due to internal forces compensate. The shear force gives
the fluid torque
\begin{equation} \label{a4}
{\bf T}_s =\dot{\gamma} \sum_n \, y_n  \,{\bf r}_n \times \hat{\bf x}, 
\end{equation} 
and the last term gives the total random torque
\begin{equation} \label{a5}
{\bf T}_r = \sum_n {\bf r}_n \times {\bf g}_n.
\end{equation} 

Let the orientation of the rod be represented by the unit vector $\hat{\bf n}$. Then the 
angular momentum can be written as
\begin{equation} \label{a6}
{\bf L} = I \, \hat{\bf n} \times d \hat{\bf n} /dt
\end{equation} 
where $I$ is the moment of inertia (divided by the mass of the monomers), equaling
\begin{equation} \label{a7}
I = \sum_n r^2_n = \frac{a^2 N^3}{12},
\end{equation} 
with $a$ is the distance between the monomers.
Since all positions ${\bf r}_n$ point in the direction  $\hat{\bf n}$, the shear torque is  
likewise
\begin{equation} \label{a8}
{\bf T}_s =\dot{\gamma} \, I \, (\hat{\bf n} \cdot \hat{\bf y}) \,  \hat{\bf n} \times \hat{\bf x}.
\end{equation} 
The correlation function between the components of the random torque follows from that
between the random forces, given by (\ref{a2}), as
\begin{equation} \label{a9}
\langle T^\alpha_r (t) \, T^\beta_r (t' ) \rangle = (2 \, k_B T \, I /\xi)  \, \delta^{\alpha,\beta} \, \delta (t-t').
\end{equation}
We have used here that the correlation tensor between the components is isotropic. 
So we may take components in the direction of $\hat{\bf n}$ and two orthogonal ones 
in the plane perpendicular to $\hat{\bf n}$. Then the non-vanishing correlations are 
multiplied by $r^2_n$ and the summation over $n$ again results in the moment of 
inertia $I$.

We get an equation of motion for $\hat{\bf n}$ by dividing the summed equations 
(\ref{a1}) by $I$
\begin{equation} \label{a10}
\hat{\bf n} \times d \hat{\bf n} /dt = \dot{\gamma} \, (\hat{\bf n} \cdot \hat{\bf y}) \,  
\hat{\bf n} \times \hat{\bf x} + {\bf T}_r/I.
\end{equation}
Clearly only components in the plane perpendicular to $\hat{\bf n}$ matter. 

We make this equation dimensionless by expressing time in terms of the rotational
diffusion time $1/(2 D_r)$, where $D_r$ is the rotational diffusion coefficient
\begin{equation} \label{a11}
D_r = \frac{k_B T}{I \xi}.
\end{equation} 
Thus we introduce the dimensionless (reduced) time variable $\tau$
\begin{equation} \label{a12}
\tau = 2 D_r t
\end{equation} 
and write (\ref{a10}) as
\begin{equation} \label{a13}
\hat{\bf n} \times d \hat{\bf n} /d \tau = W \, (\hat{\bf n} \cdot \hat{\bf y}) \,  
\hat{\bf n} \times \hat{\bf x} + {\bf t}_r,
\end{equation}
with the Weissenberg number $W$ as dimensionless measure for the shear rate
\begin{equation} \label{a14}
W = \frac{\dot{\gamma}}{2 D_r}
\end{equation} 
and ${\bf t}_r$ the reduced random torque 
\begin{equation} \label{a15}
{\bf t}_r = \frac{{\bf T}_r} {2  D_r I},
\end{equation} 
which gives the spectrum of correlation 
\begin{equation} \label{a16}
\langle t^\alpha_r (t) \, t^\beta_r (t' ) \rangle = \delta^{\alpha,\beta} \, \delta (\tau - \tau').
\end{equation}
The correlations now have a magnitude 1. As the random torques give rise to orientational
diffusion, they do this with these time units, with the orientational diffusion coefficient $1/2$.

Finally we convert (\ref{a10}) to an equation directly for the derivative of the orientation
by taking the outer product with $\hat{\bf n}$. Using that $\hat{\bf n}$ and its derivative 
are perpendicular and the general property
\begin{equation} \label{a17} 
\bf a \times (\bf b \times \bf c) = (\bf a \cdot \bf c) \,  \bf b -   (\bf a \cdot \bf b) \,  \bf c,
\end{equation} 
the equation gets the form  
\begin{equation} \label{a18}
d \hat{\bf n} /d \tau = {\bf f}_s +  \hat{\bf n} \times {\bf t}_r,
\end{equation}
where we have introduced the abbreviation
\begin{equation} \label{a19}
{\bf f}_s = W \,  \hat{\bf n} \cdot \hat{\bf y}) \,  [\hat{\bf x} - (\hat{\bf n} \cdot \hat{\bf x}) \, \,  \hat{\bf n} ], 
\end{equation} 
for the  shearing force. 
Note that the combination on the right hand side of (\ref{a19})
is a vector tangent on the unit sphere.  

\section{The probability equation}

\begin{figure}[h] 
\includegraphics[width=\linewidth]{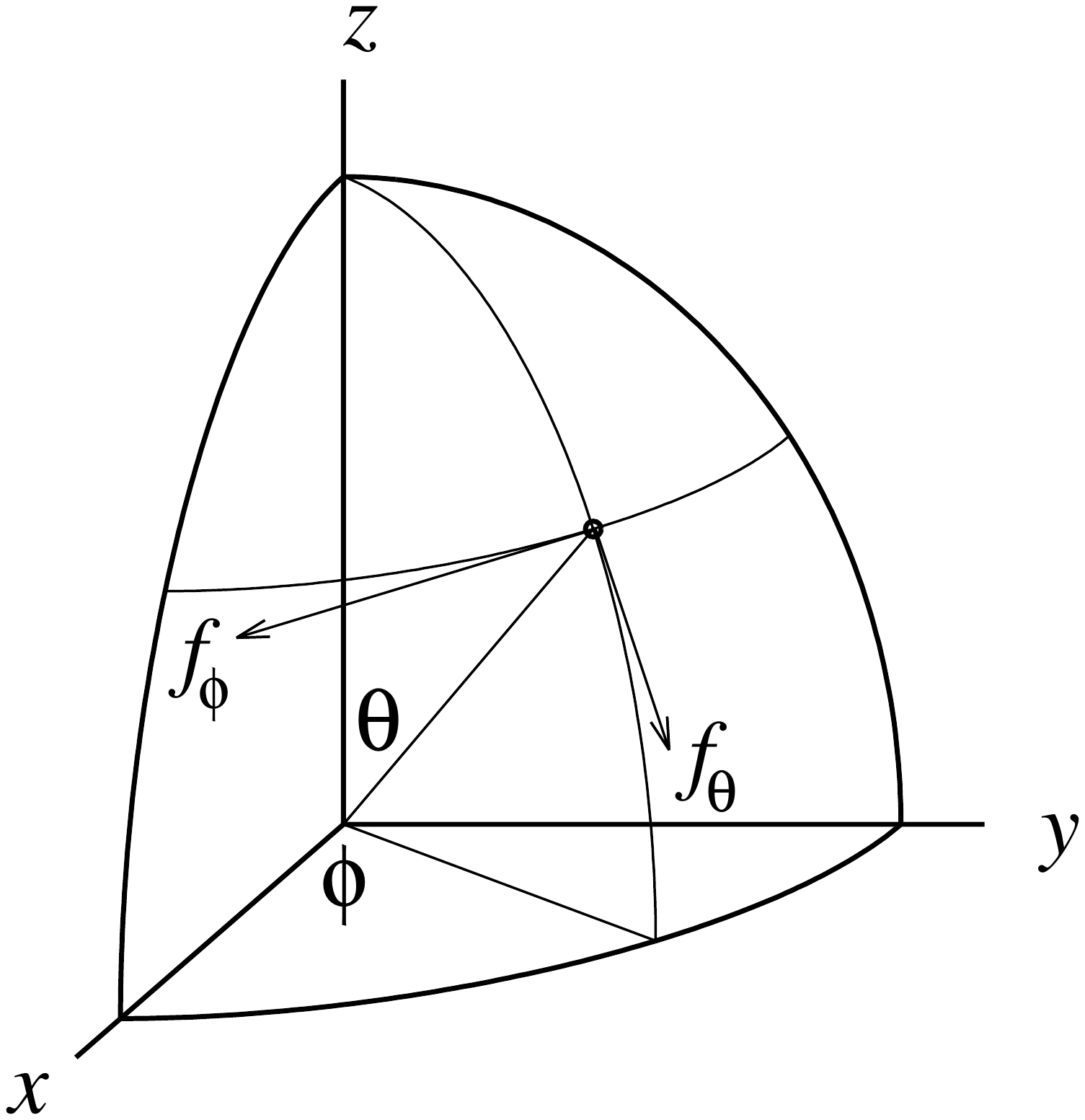}
\caption{Polar angles $\theta$ and $\phi$ and force components $f_\theta$ and $f_\phi$ of the 
force $\bf f_s$ that points in the $x$ direction with strength $f_x=W n_y (1-n^2_x)$.}\label{vec}
\end{figure}

The equations of the previous section are coordinate free. For the formulation of the
equation for the probability distribution it is convenient to use
polar coordinates, $r, \theta, \phi$. The equations of motion in terms of the polar 
coordinates follow from the geometry on the unit sphere of $\hat{\bf n}$. The $\theta$
and $\phi$ component of the shear force are given by (See Fig.~\ref{vec})

\begin{equation} \label{b1}
\left\{ \begin{array}{rcl} 
f_\theta  & = & W \sin \theta \cos \theta \sin \phi \cos \phi, \\*[4mm]
f_\phi & = & - W \sin \theta \sin^2 \phi.
\end{array} \right.
\end{equation} 
An algebraic derivation is given in Appendix \ref{force}.
As the correlation of the random vector is isotropic, we may take components in any 
coordinate system, in particular taking a torque $t_\theta$ in the $\theta$ direction and
$t_\phi$ in the $\phi$ direction. This leads to the equations 
\begin{equation} \label{b2}
\left\{ \begin{array}{rcl}
d \theta / d \tau  & = & f_\theta+ A \, t_\theta, \\*[4mm]
\sin \theta \, d \phi / d \tau    &  = &  f_\phi + A \, t_\phi.
\end{array} \right.
\end{equation} 
Here $A$ is the amplitude of the random torque satisfying equation (\ref{a15}). For a finite
timestep $\delta \tau$, the value of $A=1/\sqrt{\delta \tau}$.
We keep the $\sin \theta$ in front of the rate of change of the $\phi$ coordinate in order to 
keep the equations regular near the poles $\theta =0 $ and $\theta = \pi$.

From here the road to the Fokker-Planck equation for the probability density  
$P(\theta, \phi)$ is fairly direct. The probability distribution develops via the currents
$J_\theta$ and $J_\phi$ reading
\begin{equation} \label{b3} 
\left\{ \begin{array}{rcl}
J_\theta  & = & \displaystyle  f_\theta \, P - \frac{1}{2}  \frac{\partial P}{\partial \theta} ,\\*[4mm]
J_\phi & = & \displaystyle f_\phi  \, P - \frac{1}{2 \sin \theta}
\frac{\partial P }{\partial \phi}.
\end{array} \right.
\end{equation}  
The first term is the current due to the shear force and the second term represents the diffusive
contribution due to the random forces. As the correlation between the random torques has a unit
amplitude in our scaling, the diffusion coefficient equals 1/2.

With the currents we can write down the evolution of the probability distribution by making up
the balance between the outflow and inflow in an area  element on the sphere,
cut out but the lines of constant $\theta$ and $\phi$. The horizontal
edges are line elements of constant $\theta$ and $\theta + d \theta$ and of length 
$\sin \theta \, d \phi$ and $\sin (\theta + d \theta) \, d \phi$. The  vertical edges run at constant
$\phi$ and $ \phi + d \phi$ and have the length $d \theta$. The growth of 
$P(\theta, \phi)$ inside the area element is given by
\begin{equation} \label{b4}
\frac {\partial P(\theta, \phi, \tau)}{\partial \tau} \sin \theta \, d \theta \, d \phi.
\end{equation} 
The increase is due to difference of the flows through the horizontal and vertical
edges. The net increase through the horizontal edges  is the difference between the in-flow through
the top edge and the out-flow through the bottom edge.
\begin{equation} \label{b5}
J_\theta (\theta, \phi) \sin \theta d \phi - J_\theta (\theta + d \theta, \phi) \sin(\theta +d \theta) d \phi=
- \left( \frac{\partial J_\theta}{\partial \theta} \sin \theta + J_\theta  \cos \theta \right) d \theta d \phi
\end{equation} 
The net flow through the vertical edges equals
\begin{equation} \label{b6}
J_\phi (\theta, \phi) d \theta - J_\phi (\theta, \phi + d \phi) d \theta =
- \frac{ \partial J_\phi}{\partial \phi} d \theta d \phi.
\end{equation} 
So the balance between increase and net out-flow gives the Fokker-Planck equation
\begin{equation} \label{b7}
\frac{\partial P}{\partial \tau} = - \frac{\partial J_\theta}{\partial \theta} - 
\frac{\cos \theta}{\sin \theta} J_\theta - \frac{1}{\sin \theta} \frac{\partial J_\phi}{\partial \phi}.
\end{equation} 

We collect now the various contributions. The diffusive terms involve
\begin{equation} \label{b8}
 \Delta_{\theta, \phi} P = \left[\frac{\partial^2}{\partial \theta^2} + 
\frac{\cos \theta}{\sin \theta} \frac{\partial }{\partial \theta} + 
\frac{1}{\sin^2 \theta} \frac{\partial^2}{\partial \phi^2} \right] P
\end{equation}  
The operator on the right hand side is the angular part of the Laplacian. 
The remaining terms are collected in the operator $\cal S$ acting on the distribution
\begin{equation} \label{b9}
W {\cal S} \, P = \left[ \frac{\partial }{\partial \theta} \, f_\theta  +  \frac{1}{\sin \theta}
\frac{\partial}{\partial \phi} \, f_\phi + \frac{\cos \theta}{\sin \theta} \, 
f_\theta \right] P
\end{equation} 
So the equation for the probability density becomes
\begin{equation} \label{b10}
\frac {P(\theta, \phi, \tau)}{\partial \tau} =\frac{1}{2} \Delta_{\theta, \phi} P (\theta, \phi, \tau)- 
W \, {\cal S} \, P (\theta, \phi, \tau)
\end{equation} 
Using the expressions (\ref{b1}) for the shear forces we get in detail
\begin{equation} \label{b11}
 2 {\cal S} \, P =\sin \theta \cos \theta \sin (2\phi) \, \frac{\partial P}{\partial \theta} -
[1- \cos (2 \phi)]  \, \frac{\partial P}{\partial \phi} - 3\, \sin^2 \theta \sin (2\phi) \, P 
\end{equation}  
Equation (\ref{b10}) is the same as the one derived by Burgers (with his $\phi$ replacing our
$\pi/2-\phi$). Note that (\ref{b11}) displays the invariance under the interchange  
$(\theta, \phi) \leftrightarrow (\pi  - \theta, \pi + \phi)$.

If we put $W=0$ the equation becomes completely soluble. The modes are the spherical 
harmonics decaying exponentially in time
\begin{equation} \label{b12}
P_{l,m} (\tau) = P_{l,m} (0) \exp [- \tau  \, l (l+1)/2].
\end{equation} 
The mode $l=0$ is the stationary state and the slowest decaying mode $l=1$ decays with 
the coefficient 1, showing that (\ref{a12}) was the correct definition of the reduced time. 

\section{The Stationary State Distribution}\label{statstate}

Burgers has expanded the solution of the stationary state Fokker-Planck equation 
in powers of the Weissenberg number $W$. The stationary state follows as the solution of
\begin{equation} \label{c1}
\Delta_{\theta, \phi} \, P(\theta, \phi)  = 2 W {\cal S} \,  P(\theta, \phi).
\end{equation} 
The solution depends only on the parameter $W$. The zeroth order solution (for $W=0$)
is a constant, which we may normalize to 1. Later on we divide the probability $P$ by $4 \pi$
in order to have it normalized on the sphere. The solution thus reads
\begin{equation} \label{c2}
P (\theta, \phi) =1 + \sum_{n=1} W^n P_n (\theta, \phi),
\end{equation} 
where the $P_n$ successively follow from the equations
\begin{equation} \label{c3} 
\Delta_{\theta, \phi} \, P_n (\theta, \phi)  = 2{\cal S} \,  P_{n-1} (\theta, \phi).
\end{equation} 
By straightforward application of the operator $\cal S$ and solution of (\ref{c2}) we find
for the first term
\begin{equation} \label{c4} 
P_1 (\theta, \phi) = \frac{1}{2} \sin^2 \theta \sin 2 \phi
\end{equation} 
and for the second term
\begin{equation} \label{c5}
P_2  (\theta, \phi) = -\frac{1}{30} + \frac{1}{16} \sin^4 \theta + \frac{1}{6} \sin^2 \theta
\cos 2 \phi - \frac{1}{16} \sin^4 \theta \cos (4 \phi).
\end{equation} 
Apart from the constant $-1/30$ this agrees with the expressions given by Burgers.
Without this term the normalization of the probability distribution would not give 1.
Continuation of this process by hand gets quite involved and prone to errors. It is not
difficult to generate the terms in the expansion systematically. We have generated the 
power series and found that it does not converge beyond $W=1$, such that the use of the power
series is limited to small values of $W$.

In section \ref{statsol} we discuss methods of solution which are more powerful than the series
expansion in $W$. In this section we continue with the discussion of the flow properties in the 
stationary state, which are interesting since the currents $J_\theta$ and $J_\phi$ do not vanish. 
So there is a constant flow of probability in the solution for $P(\theta,\phi)$. 
In order to find the flow pattern we express the currents in terms of reduced flow velocities 
$v_\theta$ and $v_\phi$.
\begin{equation} \label{c6}
J_\theta = P \,\frac{d \theta}{d \tau} = P \,v_\theta, \quad \quad 
J_\phi = P \sin \theta \,\frac{ d \phi}{d \tau} = P \sin \theta \, v_\phi,
\end{equation} 
which leads, using (\ref{b3}), to the expressions for the flow velocities, 
\begin{equation} \label{c7}
\left\{ \begin{array}{rcl}
v_\theta &= & \displaystyle f_\theta- \frac{1}{2} \frac{\partial \log P}
{\partial \theta},\\*[4mm]
v_\phi &=  &\displaystyle \frac{f_\phi}{\sin \theta} - \frac{1}{2 \sin^2 \theta} 
\frac{\partial \log P}{\partial \phi},
\end{array} \right.
\end{equation} 
where $f_\theta$ and $f_\phi$ are given in (\ref{b1}). The two velocities
yield orbits in the $(\theta, \phi)$ plane, following from integrating the equations
\begin{equation} \label{c8}
\left\{ \begin{array}{rcl}
\displaystyle \frac{d \theta (\tau)}{d \tau}  & = & v_\theta (\theta(\tau), \phi(\tau)) \\*[4mm]
\displaystyle \frac{d \phi (\tau)}{d \tau} & = & v_\phi (\theta(\tau), \phi(\tau)).
\end{array} \right.
\end{equation}
The integration of the equation (\ref{c6}) must not be confused with that of 
the equations (\ref{b2}). 
The latter are a realization of a stochastic process for any initial condition. 
In (\ref{c8}) the stochastic force is replaced by the average diffusive velocity in the
stationary state (the gradient of the probability). Note that the shear part of the 
velocity $v_\phi$ vanishes for $\phi=0$. 
So the shear force stops at $\phi=0$,  where the rod is aligned with the $x$ axis. At that
point the average diffusive velocity is non-zero and pushes the rod over this dead point.
(In the Langevin equation there is no inertia which usually overcomes stagnation.) 

Rather than in this parametric form, we may obtain the orbits from direct integration of
\begin{equation} \label{c9}
\frac{d \theta}{d \phi} = \frac{v_\theta}{v_\phi}.
\end{equation} 
Since the orbits are periodic in $\phi$ we may start them all for $\phi_0 =0$ and for any
$\theta_0$ in the interval $0 \leq \theta_0 \leq \pi$. This yields $\theta(\phi; \theta_0)$ as a 
function $\phi$ and parametrically depending on $\theta_0$. For every orbit there is a 
reduced period $\tau_p(\theta_0)$ for which the increment in $\phi$ equals $2 \pi$. 
\begin{equation} \label{c10}
\tau_p(\theta_0) = \int^{2 \pi}_0 \frac{d \phi  }{|v_\phi |}.
\end{equation} 
$\theta$ will have returned to its initial position, since its derivative is a periodic function of 
$\phi$ and $\theta$ is bounded. The probability on the flow lines in the interval $d \theta_0$
around $\theta_0$ is given by $J_\phi ( \theta _0, 0) \tau_p (\theta _0)$, since $J_\phi$
gives the flow density of the rods passing at $\phi=0$ and $T (\theta_0)$ is the time it
takes for them to return to $\phi=0$. So the average reduced period $\langle \tau_p \rangle$ is then given by
\begin{equation} \label{c11}
\langle \tau_p \rangle = \int^\pi_0  d \theta_0 \, J_\phi ( \theta _0, 0) \, \tau_p^2 (\theta _0).
\end{equation} 
It is interesting to note that there exists a simple formula for the average reduced frequency $\nu$ 
of tumbling, which is the average of the inverse of $\tau_p$
\begin{equation} \label{c12}
\langle \nu \rangle = \langle 1/\tau_p \rangle = \int^\pi_0  d \theta \, J_\phi ( \theta, 0)
=\int^\pi_0  \sin \theta \, d \theta \, P(\theta ,0) \, v_\phi (\theta, 0),
\end{equation} 
showing that one does not need to in integrate the equations (\ref{c8}) in order to obtain 
$\langle\nu \rangle$. As there is a distribution over the flow lines, the two averages 
(\ref{c11}) and (\ref{c12}) are not each others inverse.

\section{The planar problem}\label{circle}

The solution of eq. (\ref{c1}) is complicated since it is defined on a sphere with
curvilinear coordinates $\theta$ and $\phi$. For illustration we first give the
solution of the problem in case that the motion would be confined to the equator of the sphere,
since this problem admits a complete analytic solution.
If the rod is restricted to the $X,Y$ plane equation (\ref{c1}) there is only a current in the
$\phi$ direction (which equals the expression (\ref{b9}) along the equator $\theta = \pi /2$)
\begin{equation} \label{d1}
J_\phi (\phi) = - W \sin^2 \phi \, P(\phi) - \frac{1}{2} \frac{d P(\phi)}{d \phi}.
\end{equation} 
Conservation of probability in the stationary state implies that the divergence of the
current vanishes
\begin{equation} \label{d2}
\frac{d J(\phi)}{d \phi} = 0, \quad \quad {\rm or} \quad \quad J(\phi) =-J_0.
\end{equation}
As the current is in the negative $\phi$ direction, we have put a minus sign before the
constant $J_0$ (which therefore obtains a positive value).
Inserting (\ref{d1}) into (\ref{d2}) gives a soluble equation for the probability distribution
with the solution 
\begin{equation} \label{d3}
P(\phi) = 2 J_0 \, u(\phi) \left( \int^{\phi}_{-\pi/2} \, \frac{d \phi'}{u(\phi')} +c_0\right),
\end{equation} 
where $u(\phi)$ defined as
\begin{equation} \label{d4}
u (\phi) = \exp\left(- W\left[\phi-\frac{1}{2} \sin (2 \phi)\right] \right).
\end{equation} 
The solution contains two constants $J_0$ and $c_0$. The latter is determined by the 
requirement that $P (\phi)$ is periodic modulo $\pi$
\begin{equation} \label{d5}
P(-\pi/2) = P(\pi/2).
\end{equation} 
This gives for $c_0$ the value
\begin{equation} \label{d6}
c_0 = \frac{\exp(-W \pi)}{1 - \exp(-W \pi)} \int^{\pi/2}_{-\pi/2} \, \frac{d \phi}{u(\phi)} 
\end{equation} 
The current $J_0$ follows from the normalization
\begin{equation} \label{d7}
\int^{\pi/2}_{-\pi/2} P (\phi) d \phi = \frac{1}{2}.
\end{equation}  
For $W=0$ the profile $P (\phi)$ is a constant $P(\phi)= 1/(2 \pi)$ and 
the current vanishes $J_0=0$. For small $W$ we may expand the integrals in powers of $W$.
To first order in $W$ we find for the constants
\begin{equation} \label{d8}
c_0 =1/W + \cdots . , \quad \quad \quad J_0 = W/(4 \pi) + \cdots .
\end{equation} 
For large $W$ the profile develops a peak for a positive value of $\phi$ near $\phi=0$. 

We get  the behaviour for asymptotically large $W$ by making the substitution 
\begin{equation} \label{d9}
\phi =y  \, W^{-1/3}   
\end{equation} 
and keeping only the largest terms in $W$. It gives the asymptotic profile
\begin{equation} \label{d10}
P(\phi) = \frac{2 J_0}{ W^{1/3}}\,  p(y),
\end{equation} 
with the function $p(y)$ given by
\begin{equation} \label{d11}
p(y) = \exp(-2 y^3/3) \int^y_{-\infty} \exp(2 y'^3/3) \, d y'.
\end{equation}  
From the normalization we now find for $J_0$ the expression
\begin{equation} \label{d12}
\frac{W^{2/3}}{2 J_0} = \int^\infty_{-\infty} dy \exp(-2 y^3/3) \int^y_{-\infty} \exp(2 y'^3/3) \, d y'.
\end{equation} 
With the value for the integral we find for $J_0$
\begin{equation} \label{d13}
J_0 = 0.07975 \, W^{2/3}.
\end{equation} 
The curve for the current $J_0$ can be found in Fig.~\ref{flowas}.
The current is very well approximated for the whole regime of $W$ values 
by the cross-over formula (\ref{g5}) with $c=0.9987$.

\section{The Stationary State Solution} \label{statsol}

The partial differential equation can be solved by making a grid on the unit sphere and 
replacing the derivatives by differences. The first step is to form a grid that is relatively
uniform and the second step is to replace the derivatives by weighted sums over the neighborhood
of the points. The boundary conditions are periodicity in $\phi$ modulo $\pi$
and symmetry between the northern and southern hemisphere.  

An alternative numerical solution expresses the probability distribution in terms of a set of 
suitable basis functions. The spherical harmonics would be such a choice, but numerically
it is a bit easier to work with the following equivalent set.  First we split $P(\theta, \phi)$ 
in an even and odd part with respect to the $\phi$ dependence.
\begin{equation} \label{e1}
P(\theta, \phi) = P_e (\theta, \phi)+P_o (\theta, \phi).
\end{equation} 
Then we express the functions as the series
\begin{equation} \label{e2}
\left\{ \begin{array}{rcl}
P_e (\theta, \phi) =\displaystyle \sum_{0 \leq m \leq k} P^{k,m}_e \sin^{2k} \theta \, \cos(2 m \phi), \\*[4mm]
P_o (\theta, \phi) =\displaystyle \sum_{1 \leq m \leq k} P^{k,m}_e \sin^{2k} \theta \, \sin(2 m \phi).
\end{array} \right.
\end{equation}
\begin{figure}[h] 
\includegraphics[width=1.2\linewidth]{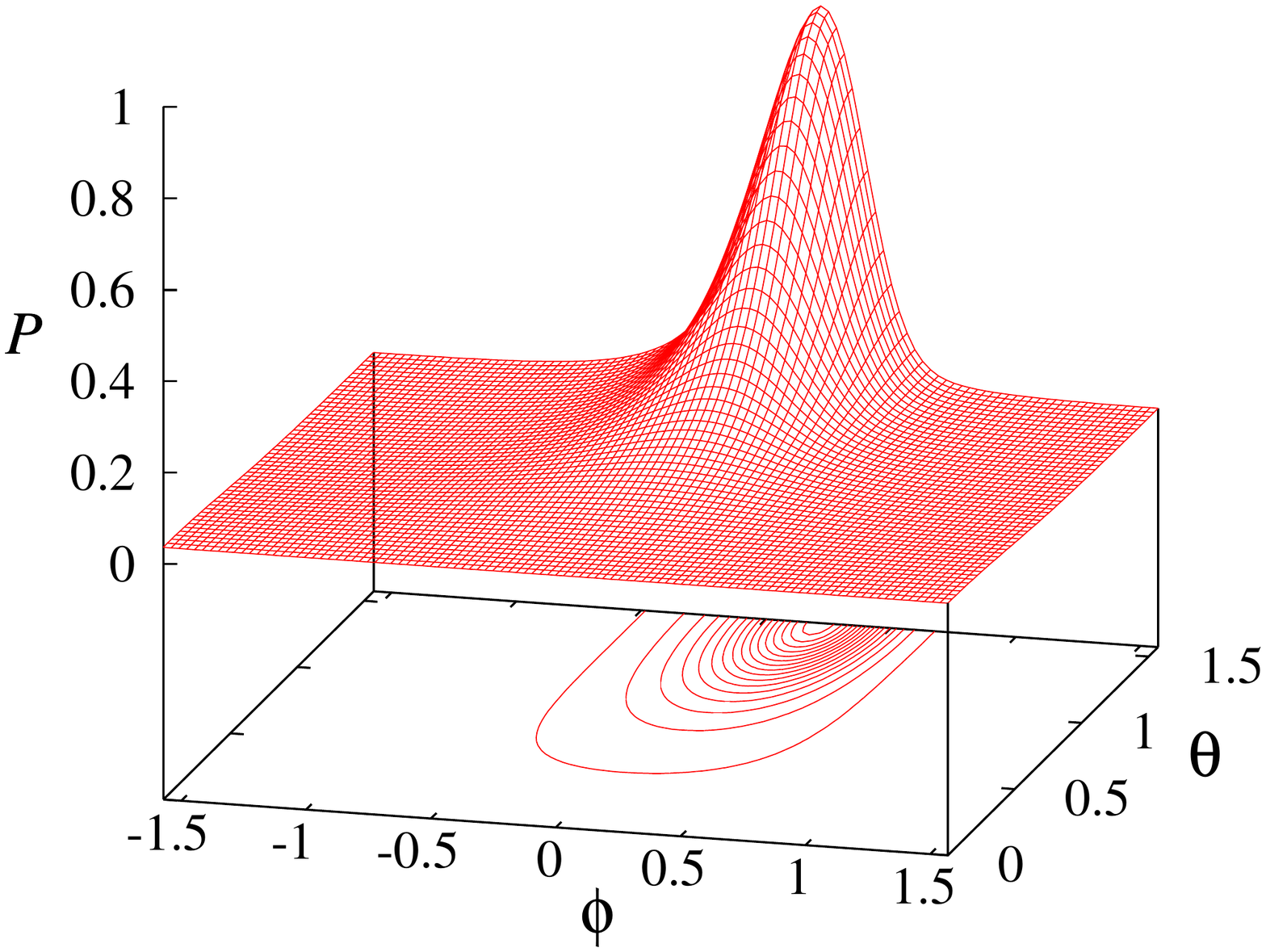}
\caption{The probability distribution for $W=30$ as a function of $\phi$ and $\theta$,
together with the contours of equal probability.}\label{prob8}
\end{figure}
The functions in the expansion have the property that $\Delta_{\theta,\phi}$ or $\cal S$ 
acting  on one of these functions results into a linear combination of these type functions. 
$\Delta_{\theta,\phi}$ turns an even function into a combination of even functions and $\cal S$
transforms an even function into a set of odd functions and vice versa.
The functions are invariant for the symmetry operations $ \theta  \leftrightarrow \pi  - \theta$ 
and $ \phi \leftrightarrow \pi + \phi $. The spherical harmonics, which are
invariant under this symmetry operation, can be expressed in terms of these functions. Since
the spherical harmonics form a complete set, the basis of (\ref{e2}) forms also a complete set.
For a numerical calculation we have to truncate the basis, say restricting the $k$ to $k<K$.
We then have $K(K+1)/2$ even functions and $K(K-1)/2$ odd functions, together a basis of
size $K^2$. We found it effective to optimize the expression
\begin{equation} \label{e3}
R = \int \sin \theta \, d \theta \, d \phi \, \left[ (\Delta - 2 {\cal S}) P (\theta, \phi) \right]^2 - 
\lambda \left[ \int \sin \theta \, d \theta \, d \phi \, P (\theta, \phi) -1 \right] 
\end{equation} 
The second term, involving the Lagrange multiplier $\lambda$,  guarantees that the optimal $P$
is normalized. $R$ is a quadratic expression in the coefficients $P_e^{k,m}$ and $P_o^{k,m}$. So
the optimal solution follows from solving a set of linear equations. In Appendix \ref{direct} we 
give details of the solution. 

A picture of the probability distribution for $W=30$ is shown in Fig.~\ref{prob8}.
The two solution methods agree in detail for $W \leq 1$. Beyond that value the method using a
grid becomes less practical as the grid has to be taken narrower. 
The method using the basis functions works
without too many functions for $W \leq 30$. Beyond that one has to use more than 
400 basis functions. In Section \ref{asymptotic} we discuss the behavior for asymptotically large $W$.

\section{The orbits}
\begin{figure}[h] 
\includegraphics[width=\linewidth]{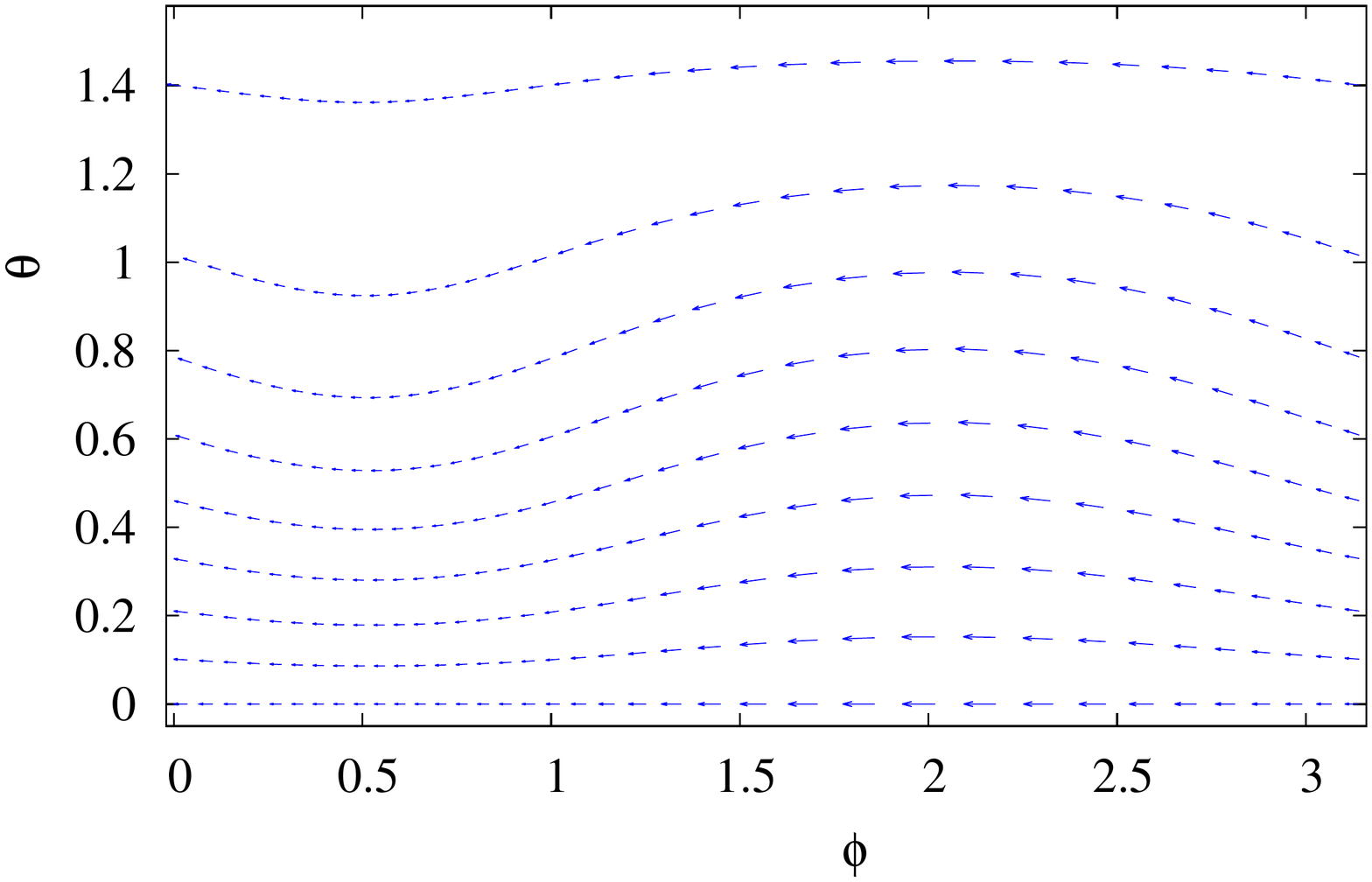}
\caption{The flow lines for $W=2$ as a function of $\phi$ and $\theta$.
The length of the arrows is proportional to the velocity.} \label{flowprob2}
\end{figure}

Before discussing the (numerical) general form of the orbits we analyze the 
low-$W$ limit. The probability $P(\theta, \phi)$ reads for small $W$ as (see 
equations (\ref{c4}) and (\ref{c5}))
\begin{equation} \label{e4}
P(\theta, \phi) = 1 -\frac{W^2}{30} + \frac{W}{2} \sin^2 \theta \left(\sin (2 \phi) + 
\frac{W}{3} \cos (2 \phi) + \frac{W}{8} \sin^2 \theta [1 - \cos (4 \phi)] \right) + 
\cdots
\end{equation} 
From this expression we find for the reduced velocities
\begin{equation} \label{e5}
\left\{\begin{array}{rcl}
v_\theta &= & \displaystyle W^2 \sin \theta \cos \theta \left(\frac{\cos (2 \phi)}{6} + 
\frac{\sin^2 \theta}{8}  [1 - \cos (4 \phi) ]\right) + \cdots \\*[4mm] 
v_\phi & = & \displaystyle - \frac{W}{2} + W^2 \left( \frac{\sin (2 \phi)}{6} - 
\frac{\sin^2 \theta}{8} \sin (4 \phi) \right) \cdots
\end{array} \right.
\end{equation} 
So in first order in $W$ ons has $v_\theta = 0$ and $v_\phi = -W/2$. This gives the lines
of constant $\theta$ as orbits with a constant flow velocity along the flow line, with the 
reduced period $\tau_p =4 \pi/W$. Translating this dimensionless time to real times we get 
$t_p = \tau_p /(2 D_r)= 4 \pi / \dot{\gamma}$. 

The expressions (\ref{e5}) are not sufficient to determine the next approximation for the flow lines 
and the tumbling time, since we would need the next order in $W$ for $v_\theta$. 
For the average reduced frequency we do not need the to evaluate the flow pattern and 
we find that there is no contribution proportional
to $W^2$
\begin{equation} \label{e6}
\langle \nu \rangle = \frac{W}{4 \pi} + {\cal O} [W^3]
\end{equation} 
In fact only odd powers in $W$ survive in the averaging process (see Appendix \ref{direct}).

Using the expansion (\ref{e1})-(\ref{e2}) one has not only an expression for the probability,
but through differentiation of the basis functions also an expression for the derivatives of 
the probability with respect to $\theta$ and $\phi$. Thus the flow pattern $v_\theta$ and $v_\phi$
can be constructed. The flow lines are
almost straight lines of constant $\theta$ for low $W$. For higher values of $W$ a flow pattern 
develops a structure, which is exhibited by a set of plots for $W=2, 10$ and 30. 
For $W=2$  a number of flow lines are given in Fig.~\ref{flowprob2}. 
Near the equator as well as near the pole the 
flow lines are again straight, but in between they undulate. 
Morover the velocity slows down near the peak in the probability. 

The next Fig.~\ref{flowprob10}  gives the picture for $W=10$. The undulation increases, in
particular, the flow lines show a deep value near the peak of the distribution. The velocities increase 
strongly with $W$ in the zone between two peaks and decrease near the peak in the probability. 
Note that in this figure the reduced velocities are devided by a factor 3 with respect to the previous 
figure. 
\begin{figure}[h] 
\includegraphics[width=\linewidth]{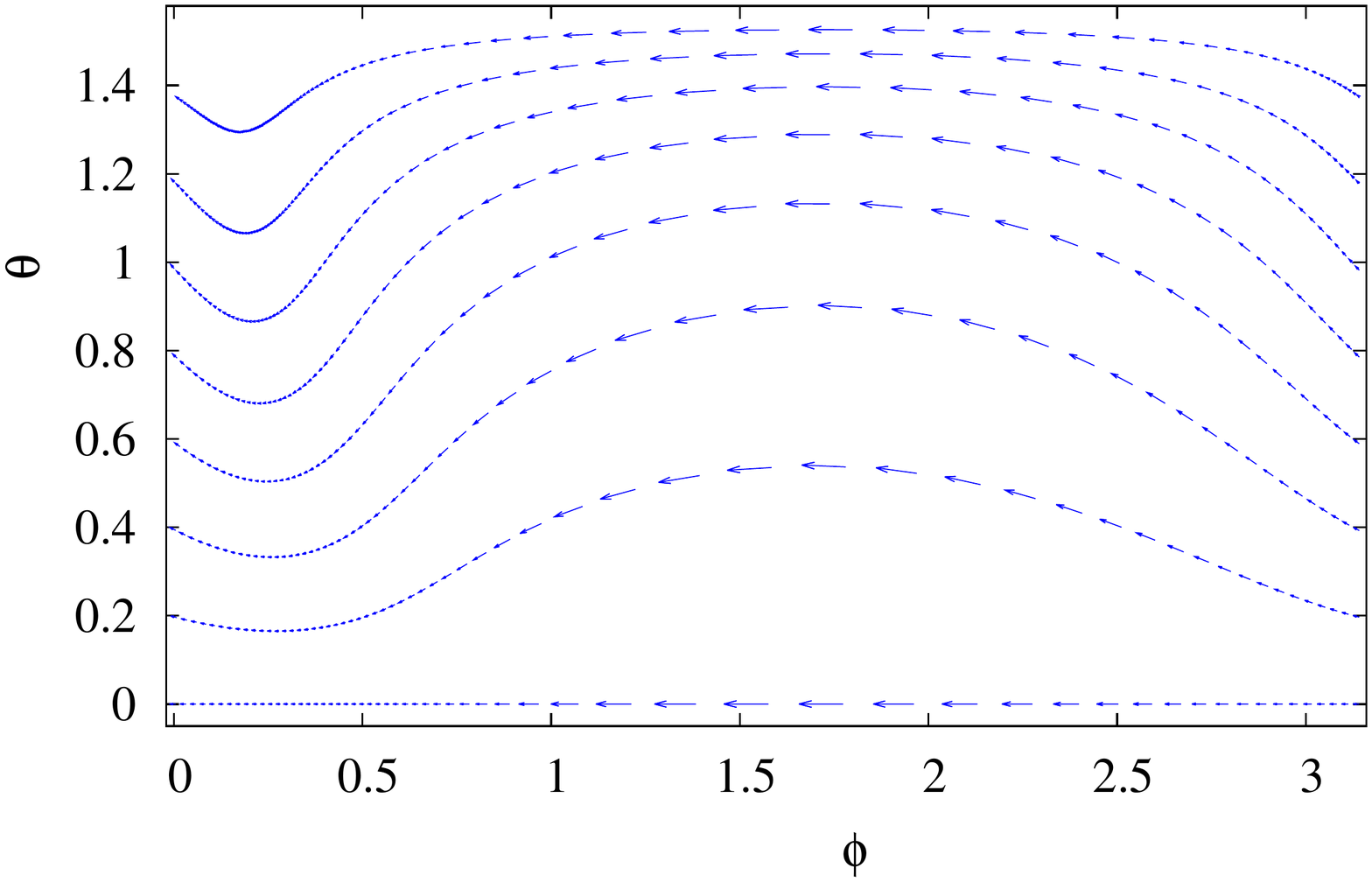}
\caption{The flow lines for $W=10$ as a function of $\phi$ and $\theta$.} \label{flowprob10}
\end{figure}

\begin{figure}[h] 
\includegraphics[width=\linewidth]{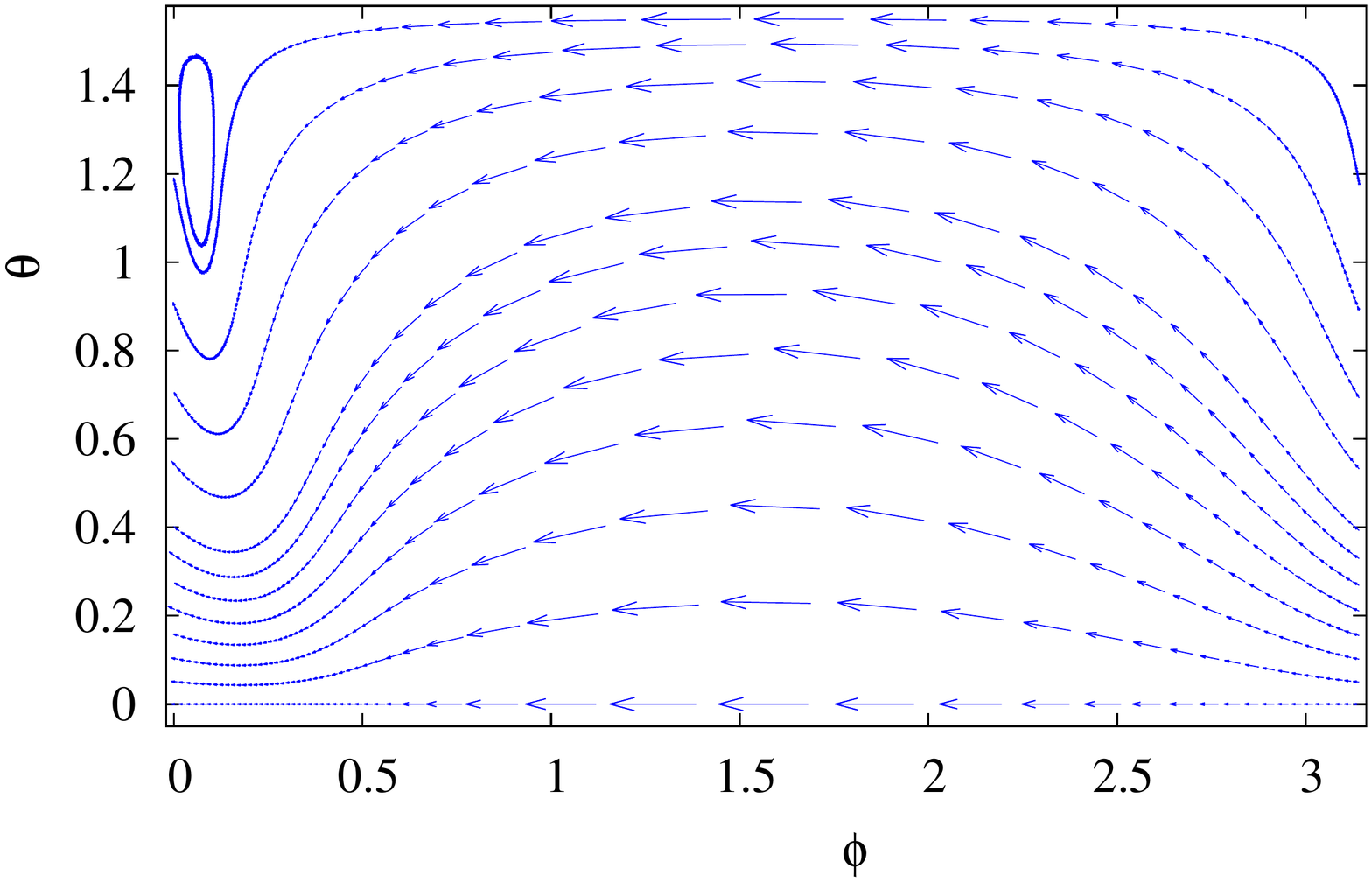}
\caption{The flow lines for $W=30$ as a function of $\phi$ and $\theta$.} \label{flowprob30}
\end{figure}

In the third Fig.~\ref{flowprob30} for $W=30$ the undulation is again stronger. 
Also the velocity pattern has  larger differences between the fast intermediate zone 
and the slow passage through the peak region.
Near the pole and small $\phi$ an eddy emerges with closed orbits not containing the pole
and not contributing to the integrated flow. This eddy nucleates around $W=20$. 

\section{Asymptotic expansion for large $W$} \label{asymptotic}

For asymptotically large $W$ we make the same substitution for the angle $\phi$ as in
the planar case and for the $\theta$ dependence we use $z=\cos \theta $ and substitute 
\begin{equation} \label{f1}
\phi =x  \, W^{-1/3}   \quad \quad {\rm and} \quad \quad z = \cos \theta = -y \, W^{-1/3}.
\end{equation}
We first rewrite the operators in terms of $z$ and $y$ and later make the substitution for $z$.
In terms of these coordinates the diffusion operator reads
\begin{equation} \label{f2}
\Delta_{\theta,\phi} = \frac{\partial}{\partial z} (1-z^2) \frac{\partial}{\partial z} + 
\frac{W^{2/3}}{(1-z^2)} \frac{\partial^2} {\partial x^2}.
\end{equation} 
The operator $\cal S$ turns into
\begin{equation} \label{f3}
2 {\cal S} =-z (1 - z^2) \sin (2 x W^{-1/3}) \frac{\partial}{\partial z}  - 2 \sin^2 (x W^{-1/3}) W^{1/3} 
\frac{\partial}{\partial x} - 3 (1- z^2) \sin (2 x W^{-1/3}).
\end{equation} 
Next we use the substitution (\ref{f1}) for $z$ and collect the leading powers in $W$
(equaling $W^{2/3}$).
\begin{equation} \label{f4}
( \Delta - 2 W {\cal S}) \tilde{P}(x,y)  =\left( \frac{\partial^2}{\partial x^2}+ 
\frac{\partial^2}{\partial y^2} + 2 x \left[ x \frac{\partial}{\partial x}  + 
2 y \frac{\partial}{\partial y} +3  \right] \right) \tilde{P}(x,y).
\end{equation} 
In the stationary state this operator acting on $P(z,y)$ must vanish. 
In the same scaling limit the currents are given by the expression
\begin{equation} \label{f5}
\left\{ \begin{array}{rcl}
\tilde{J}_x & = & \displaystyle  -x^2  \tilde{P}- \frac{1}{2} \frac{\partial \tilde{P}}{\partial x},  \\*[4mm]
\tilde{J}_y & =& \displaystyle  -xy \tilde{P} - \frac{1}{2} \frac{\partial \tilde{P}}{\partial y}.
\end{array} \right.
\end{equation} 
Equation (\ref{f4}) is the same as the condition that the divergence of the current vanishes
\begin{equation} \label{f6}
\frac{\partial \tilde{J}_x}{\partial x} + \frac{\partial \tilde{J}_y}{\partial y}  = 0.
\end{equation} 
Having obtained the (normalized) solution $\tilde{P}$ from the equation (\ref{f4}) the solution in 
terms of $\theta$ and $\phi$ is found as
\begin{equation} \label{f7}
P(\theta, \phi) = W^{2/3} \tilde{P}(-W^{1/3} \cos \theta,W^{1/3} \phi)
\end{equation} 
Note that the currents $J_\theta$ and $J_\phi$ scale as the power $W$. From the integration
over $\theta$ we get a power $W^{-1/3}$, so the integrated current scales as $W^{2/3}$.
As the period is inversely proportional
to the current, the period scales as $W^{-2/3}=(2 D_r/\dot{\gamma})^{2/3}$.
\begin{figure}[h] 
\includegraphics[width=1.2\linewidth]{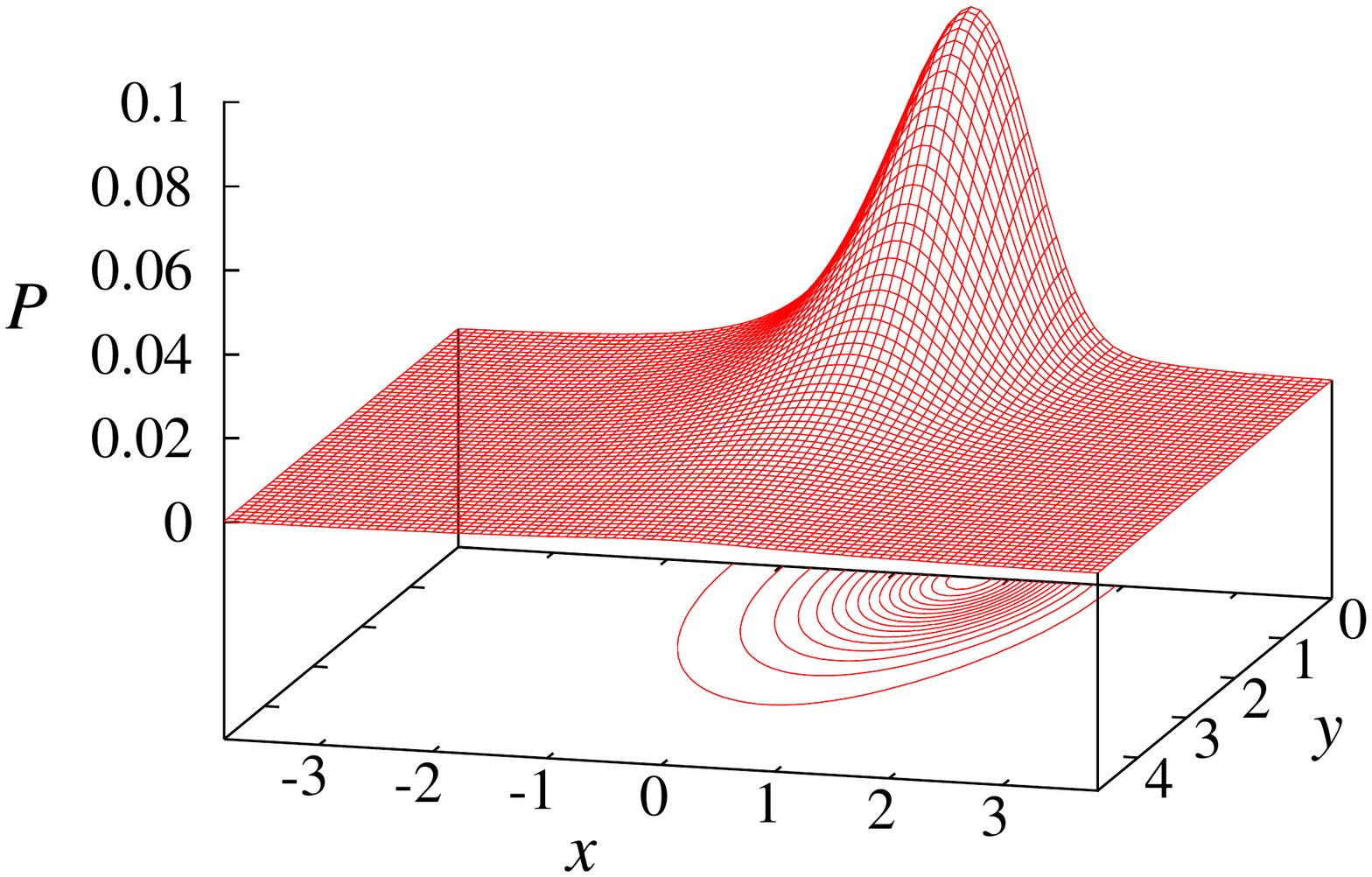}
\caption{The scaled asymptotic probability distribution.} \label{probas}
\end{figure}

\begin{figure}[h] 
\includegraphics[width=\linewidth]{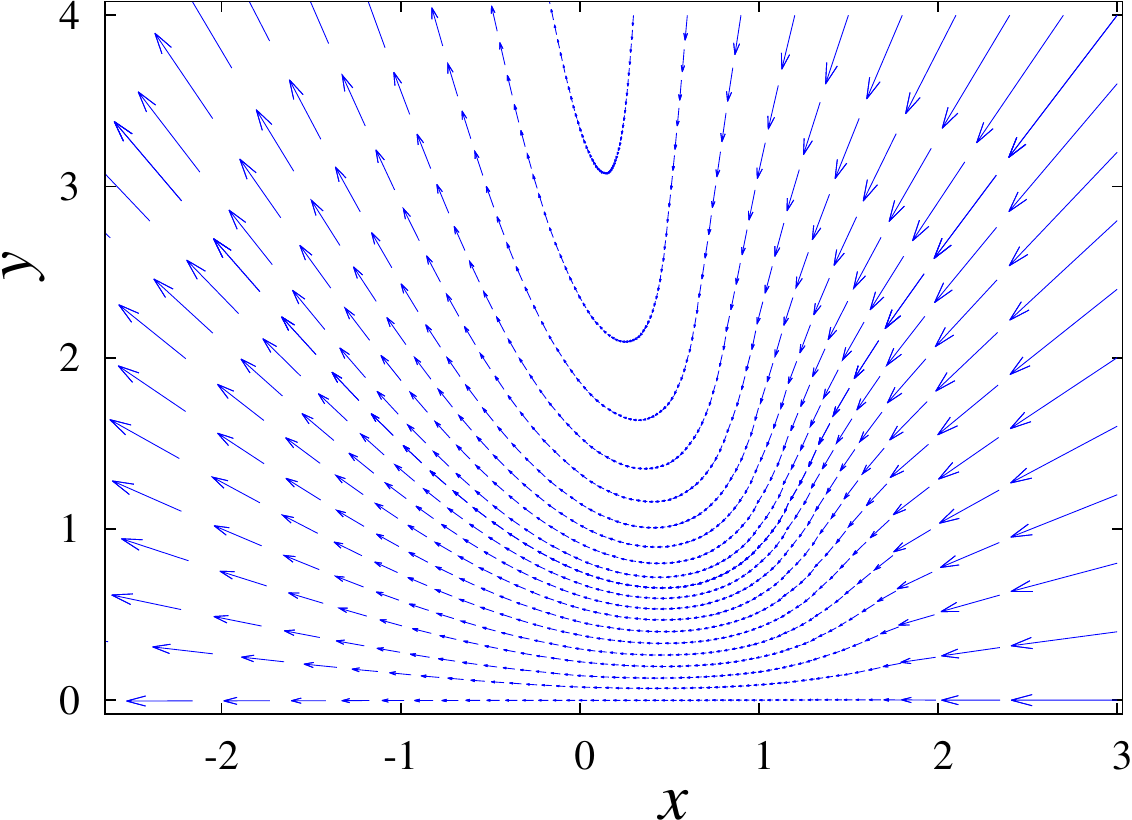}
\caption{The scaled flow pattern for asymptotically large $W$ in terms of the scaled 
variables $x$ and $y$.} \label{flowas}
\end{figure}

Since we could not find a suitable set of basis functions for the expansion of the solution,
we have taken resort to the straightforward method by solving the differential equation
through discretizing space, truncated to a finite rectangle ($-4<x<4$, $0<y<6$) 
using a 1410-point grid. As the coordinates are Cartesian the construction of the
grid and the definition of the derivatives forms no problem. The only problem are the
boundary conditions. For large values of $x$ and $y$ the diffusive terms play no role.
The remainder of the operator (\ref{f4}) shows that $\tilde{P}$ decays as $r^{-3}$ where
$r=\sqrt{x^2+y^2}$ is the distance to the center of the coordinate system. 
This condition is used to find the correction to the integrated probability
due to the outside of the rectangle, as needed for the normalization of
$\tilde{P}(x,y)$. The symmetry $y \leftrightarrow -y$ is used to define the 
boundary condition along the $y=0$ boundary of the rectangle.
The asymptotic symmetry $x \leftrightarrow -x$ is used along the three
remaining boundaries. With these boundary condition 
we found a solution for which the shape of the scaled probability $\tilde{P}$ 
is drawn in Fig.~\ref{probas}. The shape is very similar to the shape of the probability 
distribution for $W=30$. In order to get the real scale of the probability one has to multiply 
$\tilde{P}$ with $W^{2/3}$.
In Fig.~\ref{flowas} we have plotted the asymptotic flow pattern in terms of the scaled
variables $x$ and $y$. By zooming in to the region around the origin in the flow patterns
for large $W$,  as e.g. in Fig.~\ref{prob8}, one gets asymptotically this pattern. Note the
clear asymmetry in the $x$ coordinate, which remains in the scaled coordinates (but 
becomes invisible in the original variables $\theta$ and $\phi$).

\section{Discussion}

The motion of a rigid rod, immersed in a high-viscous sheared fluid-flow,  is
due to the systematic shear force and the random thermal influence of the fluid. 
An individual rod experiences a biased random motion through orientation space $(\theta, \phi)$. 
We have solved the equation for the stationary state, which gives the averages 
of this random motion over a long period of time. The flow patterns  in the stationary 
state, discussed in this paper, only reflect the average direction of motion 
for a given point in orientation space.
Time dependent correlation functions such as the probability to arrive at time $t$ at
a point $(\theta, \phi)$, after starting from a point $(\theta', \phi')$, would involve the solution
of the Fokker-Planck equation as an initial value problem.
 
The stationary equation contains, after appropriate scaling, as the only physical parameter the
Weissenberg number $W$, which is the ratio of the shear rate $\dot{\gamma}$ and the 
orientational diffusion constant $D_r$. According to (\ref{a11}) the latter is determined 
by the temperature $T$, the viscosity $\xi$ and the moment of inertia $I$. The usual flowing fluid
is water at room temperature for which we have the values 
\begin{equation} \label{g1}
k_B T = 4 \, {\rm pN \, nm}, \quad \quad \quad \xi = 2 \, \cdot 10^{-12} \, {\rm kg / s}.
\end{equation}
The moment of inertia is given by the expression (\ref{a7}), in which 
$a$ is the distance between the monomers. The value $a=0.33$ nm, which holds for dsDNA,
is a reasonable number.
$N$ is the number of monomers, which can vary from a few to numbers as high as  $10^4$  for
f-actin, still keeping the polymer fairly rigid. Using (\ref{g1}) and (\ref{g2}) in  (\ref{a11}) gives
for the inverse diffusion coefficient
\begin{equation} \label{g2}
\frac{1}{2 D_r} = \frac{I \xi}{2 k_B T} = 2.3 \cdot 10^{-12} N^3 \, {\rm s}
\end{equation} 
and the associated Weissenberg number 
\begin{equation} \label{g3}
W = \frac{\dot{\gamma}}{2 D_r} =2.3 \cdot 10^{-12} \,  N^3 \, \dot{\gamma} \, {\rm s}.
\end{equation} 
Thus by varying the shear rate and the length of  the polymer one can cover a wide range of 
Weissenberg numbers from extremely small for short polymers to fairly large for long stiff polymers.

For small $W$ the expansion of the solution in powers of $W$ suffices. The (average) period
scpfor tumbling in real time is independent of the size of the polymer and given by $4 \pi/\dot{\gamma}$,
only depending on the shear rate. For intermediate values of $W$ the solution as discussed in 
section \ref{statsol} can be used. In the reduced time $\tau$ this gives a period crossing over
from the $W^{-1}$ behavior at small $W$ to the $W^{-2/3}$ behavior at asymptotically large $W$.
The intermediate regime has an interesting flow pattern, showing the emergence of a vortex
near the pole for small values of $\phi$. For asymptotically large $W$ the period scales 
in real time scales as $D_r^{-1/3} \dot{\gamma}^{-2/3} \sim \dot{\gamma}^{-2/3} N$.

\begin{figure}[h] 
\includegraphics[width=\linewidth]{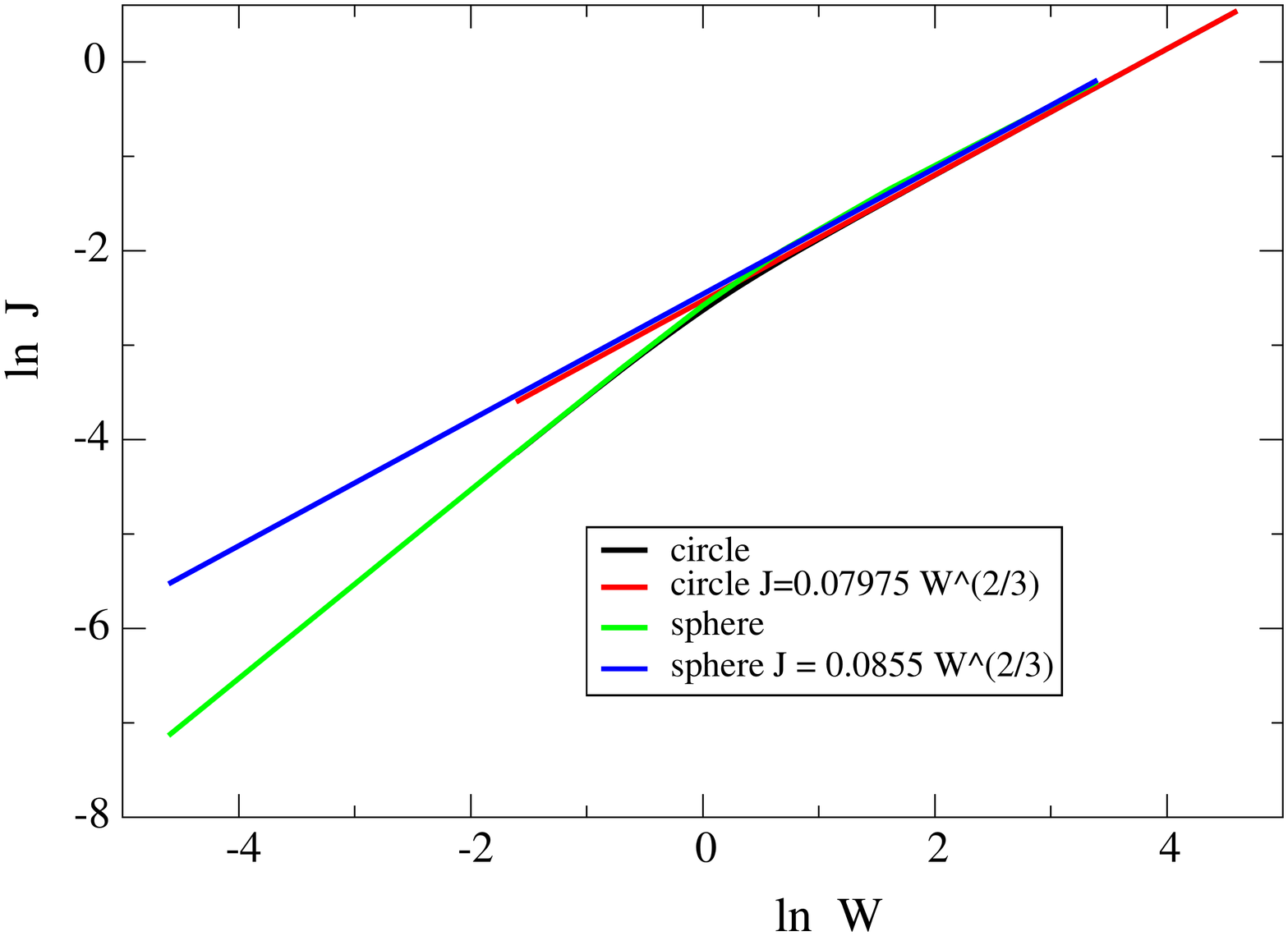}
\caption{The current as a function of the Weissenberg number} \label{current}
\end{figure}

\begin{figure}[h] 
\includegraphics[width=\linewidth]{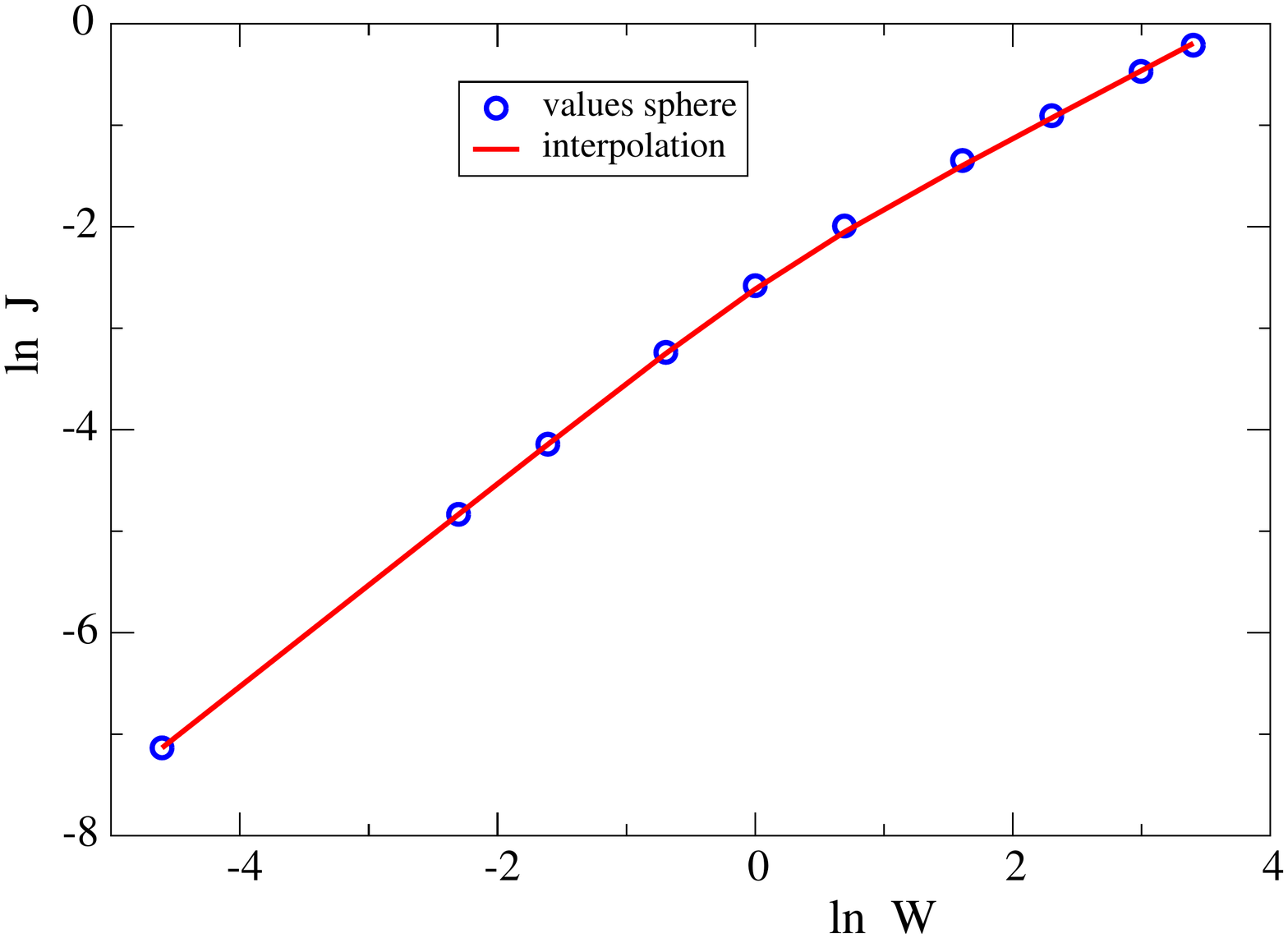}
\caption{The average current as a function of the Weissenberg number and the interpolation 
formula (\ref{g5}) for the spherical problem.} \label{interpol}
\end{figure}

There exist an easy expression for the total current $J$, which equals the average reduced tumbling
frequency $\langle \nu \rangle$ 
\begin{equation} \label{g4}
\langle \nu \rangle = J = \int^\pi_0 d \theta J_\phi (\theta, \phi), 
\end{equation} 
which is independent of $\phi$, due to conservation of probability.
The curve is very similar for the planar and  spherical problem as is indicated
in Fig.~\ref{current}.
In fact, an excellent approximation is given by the interpolation formula
\begin{equation} \label{g5}
\langle \nu \rangle \simeq \frac{W}{4 \pi (1 + c W^2)^{1/6}},
\end{equation} 
with $c$ determined from the asymptotic current. We find $c=0.650$ for the sphere and $c=0.9987$
for the circle. 
The average reduced frequency and the interpolation formula are shown in Fig.~\ref{interpol} for the 
spherical problem. 

\appendix

\section{The forces $f_\theta$ and $f_\phi$ }\label{force}

\setcounter{equation}{0}
\renewcommand{\theequation}{A\arabic{equation}}

In this appendix we list some of the relations between the polar and Cartesian coordinates.
The Cartesian components of the orientation $\hat{\bf n}$ read
\begin{equation} \label{A1}
(n_x,\, n_y,\,  n_z)=(\sin \theta \cos \phi, \,\sin \theta \sin \phi,\, \cos \theta).
\end{equation} 
The relations between the time derivatives is
\begin{equation} \label{A2}
\left\{ \begin{array}{rcl}
 \displaystyle \frac{d n_x}{d \tau} & = & \displaystyle \cos \theta \cos \phi \, \frac{ d \theta}{d \tau}
    -\sin \theta \sin \phi \frac{d \phi}{d \tau}, \\*[4mm]
 \displaystyle \frac{d n_y}{d \tau} & = & \displaystyle \cos \theta \sin \phi \, \frac{ d \theta}{d \tau}
    +\sin \theta \cos \phi \frac{d \phi}{d \tau}, \\*[4mm]
 \displaystyle \frac{d n_z}{d \tau} & = & \displaystyle -\sin \theta \, \frac{ d \theta}{d \tau}.
\end{array} \right.
\end{equation} 
The time derivative of the angle are expressed in the Cartesian components as
\begin{equation} \label{A3}
\left\{ \begin{array}{rcl}
 \displaystyle \sin \theta \, \frac{d \theta }{d \tau} & = & \displaystyle -
\frac{d n_z}{d \tau} , \\*[4mm]
 \displaystyle \sin \theta \, \frac{d \phi }{d \tau} & = & \displaystyle \cos \phi \, 
\frac{d n_y}{d \tau}-\sin \phi \, \frac{d n_x}{d \tau}.
\end{array} \right.
\end{equation} 
The shear force are given by (\ref{b1})
\begin{equation} \label{A4}
\left\{ \begin{array}{rcl}
f_x & = & W n_y \, (1 - n_x^2), \\*[2mm]
f_y & = & W n_y \, (-n_x n_y), \\*[2mm]
f_z & = & W n_y \, (-n_x n_z).
\end{array} \right.
\end{equation} 
Since 
\begin{equation} \label{A5}
n_x \, \sin \phi = n_y \, \cos \phi, 
\end{equation} 
we find for the equations without the random forces
\begin{equation} \label{A6}
\left\{ \begin{array}{rcl}
 \displaystyle \sin \theta \, \frac{d \theta }{d \tau} & = & W n_y n_x n_z,\\*[4mm]
\displaystyle \sin \theta \, \frac{d \phi }{d \tau} & = & - W \sin \phi \,  n_y.
\end{array} \right.
\end{equation} 
Inserting the values of $n_x,n_y$ and $n_z$ as given by (\ref{A1}) gives agreement with (\ref{b1}).

\section{Optimal solution of the differential equation} \label{direct}

\setcounter{equation}{0}
\renewcommand{\theequation}{B\arabic{equation}} 

The solution of the probability equation (\ref{c1}) via an expansion in suitable
basis functions requires to evaluate the action of the operators $\Delta$ and $\cal{S}$ 
on a member of the set. We define a matrix for the operator $\Delta$ as
\begin{equation} \label{B1}
 \Delta_{\theta,\phi} (\sin \theta)^{2k} \cos 2 m \phi = \sum_{k',m'} ( \sin \theta )^{2k'}  \cos (2 m' \phi) 
\langle k', m' | \Delta | k,m \rangle.
\end{equation}  
The action of Delta a basis function is relative simple:
\begin{equation} \label{B2}
\Delta \sin^{2k}  \theta \cos(2 m \phi) = [-2k(2k+1) \sin^{2k}  \theta + 4(k^2 -m^2) 
\sin^{2(k-1)} \theta ]\, \cos(2 m \phi). 
\end{equation} 
So we find the non-zero matrix elements
\begin{equation} \label{B3}
\left\{ \begin{array}{rcl}
\langle k, m |\, \Delta \,| k, m \rangle & = & - 2k(2k+1), \\*[2mm]
\langle k-1, m |\,\Delta \,| k, m \rangle & = & 4 (k^2 - m^2).
\end{array} \right.
\end{equation} 
These relations hold equally for the even $\cos (2m \phi)$ as the odd $\sin (2m \phi)$ functions.
Note that the second term vanishes for $k=m$.

Similarly a matrix, accounting for the action of $\cal{S}$ is defined as
\begin{equation} \label{B4}
 {\cal S} (\sin \theta)^{2k} \cos 2 m \phi = \sum_{k',m'} ( \sin \theta )^{2k'}  \sin (2 m' \phi) 
\langle k', m' |  {\cal S} \, | k,m \rangle.
\end{equation}  
For the operation on an odd function practically the same matrix can be used with an overall
minus sign and a few changes related to the fact that the odd basis is smaller than the even basis.
The matrix of $\cal S$ is more complicated as we have three terms in the expression. We
treat them separately. The first term yields
\begin{equation} \label{B5}
\begin{array}{c}
\displaystyle \sin \theta \cos \theta \sin(2 \phi) \frac{\partial }{\partial \theta} \sin^{2k}  \theta \cos(2 m \phi) = \\*[4mm]
k \, [\sin^{2k} \theta - \sin^{2(k+1)} \theta]\, [ \sin (2(m+1) \phi) - \sin (2(m-1) \phi)].
\end{array}
\end{equation} 
The second term gives 
\begin{equation} \label{B6}
\begin{array}{c}
\displaystyle  -[1- \cos (2 \phi)] \frac{\partial}{\partial \phi} \sin^{2k}  \theta \cos(2 m \phi) = \\*[4mm]
m \sin^{2k}  \theta \, [ 2 \sin (2 m \phi) -\sin (2(m+1) \phi) - \sin (2(m-1) \phi)]
\end{array}
\end{equation}
The third term gives
\begin{equation} \label{B7}
- 3 \sin^2 \theta \sin (2 \phi) \sin^{2k}  \theta \cos(2 m \phi) = 
- \frac{3}{2} \sin^{2(k+1)} \theta \, [ \sin (2(m+1) \phi) - \sin (2(m-1) \phi)]
\end{equation} 
So we find the following non-zero coefficients for the matrix of $ 2 {\cal S}$
\begin{equation} \label{B8}
\left\{ \begin{array}{rcl}
\langle k+1, m+1 |  \,2 {\cal S} \, | k,m \rangle & = & -(k+3/2), \\*[2mm]
\langle k+1, m-1 |  \,2 {\cal S}  \,| k,m \rangle& = & k+3/2, \\*[2mm]
\langle k, m |  \, 2{\cal S}  \,| k,m \rangle & = & 2m, \\*[2mm]
\langle k, m+1 |  \, 2{\cal S} \, | k,m \rangle & = & k-m, \\*[2mm]
\langle k, m-1 |  \,2 {\cal S} \,| k,m \rangle & = & -(k+m). \\*[2mm]
\end{array} \right.
\end{equation} 
The  transition from $m=0$ to $m=-1$ does not exist of course. Inspecting the formulas one sees 
that $\sin(2(0-1) \phi)$ means $-\sin (2 \phi)$. So the second and fifth entry have to be dropped
and the first and fourth entry doubled for $m=0$.

These relations apply to the case where the input is the an even function. An odd function gives
the same relations with a minus sign. The case $m=0$ does not occur in the odd function space.

We may now write the stationary state equation (\ref{c1}) symbolically as
\begin{equation} \label{B9}
\left( 
\begin{array}{cc}  \Delta  &-2 W {\cal S}  \\*[4mm] 
                                      2 W {\cal S} & \Delta  
\end{array} \right) 
\left(
\begin{array}{c} P_e \\*[4mm] P_o \end{array} 
\right) = 0.
\end{equation} 
This shows that the problem of finding the stationary state distribution is equivalent to the
determination of the right eigenvector belonging to the eigenvalue 0 of the matrix. The left
eigenvalue can easily be found due to conservation of probability, which is implied by the 
property
\begin{equation} \label{B10}
\int^\pi_0 \sin \theta \, d \theta \int^\pi_0 d \phi \, {\cal S} f(\theta, \phi) = 0.
\end{equation} 
The proof of this relation follows by partial integration. For $\Delta_{\theta,\phi}$ the same 
property holds. Thus for any input function the action of $\Delta$ and $\cal S$
gives a function that integrates to zero, or
\begin{equation} \label{B11}
\int^\pi_0 \sin \theta \, d \theta \int^\pi_0 d \phi \, \sum_{k',m'} ( \sin \theta )^{2k'}  \cos (2 m' \phi) 
\langle k', m' | \Delta \, | k,m \rangle =0.
\end{equation} 
Carrying out the integration, only the terms with $m'=0$  contribute and give the coefficients
\begin{equation} \label{B12}
q_k = \int^\pi_0 \sin \theta \, d \theta \, \sin^{2k} \theta.
\end{equation} 
The $q_k$ follow recursively from $q_0=2$ and 
\begin{equation} \label{B13}
q_k = \frac{2 k}{2 k+1} q_{k-1}.
\end{equation} 
Thus $q_k \delta_{m,0}$ is a left eigenvector of the matrix $\Delta$ with eigenvalue 0.
This holds also for the larger matrix  (\ref{B9}) since $\cal S$ acting on an odd function
gives only functions that integrate to zero and action on an even function gives odd functions
that integrate a fortiory to zero.

A strong test of the accuracy of the optimization is the fact that the average current 
\begin{equation} \label{B14}
\langle J_\phi \rangle = \int^{\pi}_0 d \theta \, J_\phi(\theta, \phi),
\end{equation} 
has to be independent of the angle $\phi$ due to conservation of probability. As the integral
is obtained as a series in $\cos(2 m \phi)$ and $\sin(2 m \phi)$ all the terms with $m \neq 0$ 
must vanish. This gives a set of relations between the $P_e^{k,m}$ and $P_0^{k,m}$. We have
verified that these relations are fulfilled with an accuracy that deteriorates somewhat for
large $W$. The terms with $m=0$ survive and give the average current as
\begin{equation} \label{B15}
\langle J_\phi \rangle = \frac{W}{2} \left( 1 - \sum_k P_e^{k,1} q_k \right),
\end{equation} 
where we have used the normalization 
\begin{equation} \label{B16}
1 = \sum_k P_e^{k,0} q_k.
\end{equation}

\end{document}